\documentclass[aps,pra,reprint,amssymb,amsmath,longbibliography]{revtex4-2}

\usepackage{hyperref}
\usepackage{graphicx}
\usepackage{color}
\usepackage{braket}
\usepackage{xcolor}

\begin{document}

\title{Multipartite entanglement generation in coupled microcavity arrays}
\author{Marc Bostelmann}
\author{Steffen Wilksen}
\author{Frederik Lohof}
\author{Christopher Gies}
\affiliation{Institute for Theoretical Physics and Bremen Center for Computational Material Science, University of Bremen, Bremen, Germany}
\date{\today}

\begin{abstract}
	We consider photonic arrays made from quantum emitters in optically coupled microcavities as a platform for entanglement generation.
	These offer a large degree of tunability with the possibility of site-selective optical excitation.
	Coherent pumping is considered to drive transitions between vacuum and entangled target states both in a time-dependent manner, and in a quantum bath engineering approach to create entanglement in the steady-state.
	We demonstrate a numerical scheme that allows to generalize the determination of excitation parameters to larger array sizes and different classes of entangled states.
	This study is a step towards using coupled cavity arrays as a hardware platform in novel quantum-photonic applications in quantum computing and quantum machine learning.
\end{abstract}

\maketitle

\section{Introduction}
\label{sec:intro}
Solid-state systems operating in the regime of cavity quantum electrodynamics (cQED) are being considered for many applications in optoelectronics, ranging from nanolasers \cite{deng_physics_2021} to quantum-light sources \cite{senellart_high-performance_2017} and sensors \cite{yoshie_correction_2011} up to applications in quantum reservoir computing \cite{heuser_developing_2020}.
The modification of the optical density of states due to the presence of a cavity allows for tailoring and enhance spontaneous emission, which can be used to increase brightness and efficiency in a wide range of material systems \cite{deng_physics_2021,reeves_2d_2018,romeira_purcell_2018}.
Amongst these, quantum dots (QDs) are particularly interesting for applications in the quantum technologies, as their properties can be precisely tuned to match the embedding resonator structure \cite{moczala-dusanowska_strain-tunable_2019,nowak_deterministic_2014, schmidt_deterministically_2020}.
They can function as a platform for entanglement generation \cite{delley_deterministic_2017}:
By adding photonic connections between the cavities, e.g.~by a network of waveguides \cite{van_der_sande_advances_2017} or external mirrors \cite{heuser_fabrication_2018}, fully or partially connected coupled cavity arrays (CCAs) can be realized \cite{grujic_non-equilibrium_2012,hartmann_quantum_2008,hartmann_strong_2007}.
In the spatially homogeneous case, CCAs foster collective modes that are delocalized over the whole array \cite{ruiz-rivas_spontaneous_2014}.
In these collective modes, photons can induce correlations in the electronic degrees of freedom of the quantum emitters that are located in distant cavities, perfectly combining an electronic system that is capable of hosting both classical and quantum correlations \cite{mascarenhas_laser_2016} with convenient accessibility of each individual cavity by optical excitation.
For this reason, CCAs are of particular interest for several emerging technologies, such as the deterministic generation of multipartite entanglement (MPE) \cite{sun_generation_2016} and quantum reservoir computing \cite{fujii_harnessing_2017}.
CCAs have previously been considered for the generation of cluster states \cite{sun_generation_2016}, Bell states \cite{hein_purification_2016}, and GHZ states \cite{aron_photon-mediated_2016}.
Another important class of MPE states are W states, which stand out due to their robustness against particle loss and their application in quantum communication protocols \cite{bellomo_n_2017}.
In general, MPE (as exhibited by GHZ and W states) is of great interest because of its superiority over bipartite entanglement from the viewpoint of state convertibility by local operations and classical communication (LOCC) \cite{yamasaki_multipartite_2018}.
Experimentally, W states (and other MPE states) have been generated by entangling photons \cite{eibl_experimental_2004}, superconducting qubits \cite{gong_genuine_2019}, or trapped ions \cite{haffner_scalable_2005}.
Important platforms for experimental generation of spatially distributed quantum systems are atomic qubits \cite{hucul_modular_2015,ritter_elementary_2012}, superconducting qubits \cite{daiss_quantum-logic_2021}, spins in diamonds \cite{humphreys_deterministic_2018}, spins in semiconductor quantum dots \cite{stockill_phase-tuned_2017}, and trapped ions \cite{stephenson_high-rate_2020}.
Furthermore, different theoretical schemes for generating W states have been put forward \cite{bellomo_n_2017,brus_multipartite_2005,kim_efficient_2020,song_quantum_2022, wang_efficient_2015, wu_highly_2018}.

The purpose of this work is to analyze in detail how entanglement can be generated in CCAs using optical techniques.
We provide a detailed analysis of the eigenspectrum and identify transitions into entangled target states, taking into account pump-induced energy renormalizations and symmetry considerations.

In Section~\ref{sec:CCAeigenspectrum} we introduce the system of CCAs and provide intuitive insight to its eigenstates in the presence of dissipative processes and coherent excitation, which is the foundation for the targeted excitation of entangled states using coherent excitation pulses in Section~\ref{sec:egscheme}.
The generation scheme we introduce in this section is generalized to different classes of MPE target states in Section~\ref{sec:mpe}, where we demonstrate the scalability of our approach to target states of three and four qubits.
The influence of symmetry and the possibility of generating antisymmetric states, such as phased W states, is the topic of Section~\ref{sec:symmetrysteering}.
Steady-state entanglement is a particularly relevant concept for novel quantum machine learning approaches, such as quantum reservoir computing, and is currently being investigated in the context of quantum bath engineering \cite{gramajo_efficient_2021,ullah_steady_2022, oliveira_steady-state_2022}.
Therefore, we provide additional insight both into steady-state MPE generation, and the bath engineering concept in the framework of a Bloch-Redfield approach in Section~\ref{sec:steadystate} \cite{aron_steady-state_2014}.

\section{The CCA model and its eigenspectrum}
\label{sec:CCAeigenspectrum}
\begin{figure}
	\centering
	\includegraphics{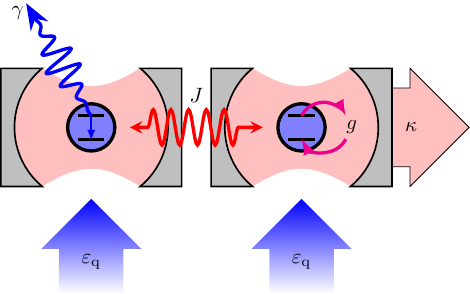}
	\caption{Two cavities coupled with interaction strength $J$.
	Each cavity contains a qubit that is driven by an external laser with amplitude $\varepsilon_\mathrm{d}$.
	The interaction between the cavities and the qubits is described by the Jaynes-Cummings model with the light-matter interaction strength $g$.
	Qubits and cavities lose energy through dissipative processes at the rates $\gamma$ and $\kappa$, respectively.}
	\label{fig:cavities}
\end{figure}

We consider systems consisting of $ N $ single-mode cavities each containing a two-level emitter (qubit) as shown in Fig.~\ref{fig:cavities}.
Adjacent cavities are coupled due to the overlap of their photonic modes \cite{hartmann_quantum_2008,moughames_three-dimensional_2020}.
The qubits are driven coherently by external lasers whose respective frequencies can be detuned from those of the qubits.
The Hamiltonian of the system is that of a driven Jaynes-Cummings-Hubbard system \cite{hartmann_strongly_2006, greentree_quantum_2006} and reads
\begin{equation}
	H = H_\mathrm{q} + H_\mathrm{c} + H_\mathrm{c,q},
	\label{eq:hamiltonian_general}
\end{equation}
where $ H_\mathrm{q} $, $ H_\mathrm{c} $, and $ H_\mathrm{c,q} $ are the qubit, cavity, and cavity-qubit coupling Hamiltonian, respectively.
These are given by (we set $ \hbar = 1 $)
\begin{align}
	H_\mathrm{q} &= \sum_{i=1}^{N} \left[ \omega_\mathrm{q} \sigma_i^{+} \sigma_i^{-} + 2\varepsilon_\mathrm{q}\cos\left(\omega_\mathrm{d}t + \phi_i\right) \left(\sigma_i^{+} + \sigma_i^{-}\right) \right],\label{eq:Hq}\\
	H_\mathrm{c} &= \sum_{i=1}^N \omega_\mathrm{c} a_i^\dagger a_i - \sum_{i,j} J_{ij} a_i^\dagger a_j,\label{eq:Hc}\\
	H_\mathrm{c,q} &= \sum_{i=1}^N g \left(a_i^\dagger \sigma_i^{-} + a_i \sigma_i^{+}\right).\label{eq:Hcq}
\end{align}
The intercavity hopping parameters $ J_{ij} $ between cavities $ i $ and $ j $ determine the network topology, and we use the convention $ J_{ii} \equiv 0 $ and $ J_{ij} = J_{ji} $ for $J_{ij}\in\mathbb{R}$.
We consider dissipative processes in terms of the Lindblad master equation,
\begin{equation}
	\frac{\mathrm{d}}{\mathrm{d}t} \rho = \mathcal{L} \rho \equiv -\mathrm{i} [H, \rho] + \sum_{n} \mathcal{D}[C_n]\rho,
	\label{eq:lindblad}
\end{equation}
with the (non-unitary) Liouvillian superoperator $\mathcal{L}$ and the Lindblad dissipators
\begin{equation}
	\mathcal{D}[C_n]\rho \equiv \frac{1}{2}\left(C_n \rho C_n^\dagger - C_n^\dagger C_n \rho + \mathrm{h.c.}\right),
\end{equation}
where $ C_n = \sqrt{\gamma_n} A_n $ are the collapse operators with an operator $ A_n $ that couples the system to an environment at rate $ \gamma_n $.
The relevant processes considered in this work are cavity decay ($C_\kappa= \sqrt{\kappa} a_i $), qubit decay ($C_\gamma= \sqrt{\gamma} \sigma_i^{-} $) and pure dephasing ($C_{\gamma_\varphi}=\sqrt{\gamma_\varphi} \sigma_i^{z} $).

It is convenient to move into the rotating frame \mbox{$ H \mapsto U \left(H - \mathrm{i}\partial_t\right) U^\dagger $}, by applying
\begin{equation}
	U = \prod_{i=1}^{N} \exp \left[\mathrm{i}\omega_\mathrm{d}t \left(\sigma_i^{+} \sigma_i^{-} + a_i^\dagger a_i\right)\right],
\end{equation}
allowing us to neglect the fast-rotating terms of the form $ \mathrm{e}^{2\mathrm{i}\omega_\mathrm{d}t}a_i^\dagger $.
As a consequence, the bare energies $ \omega_\mathrm{q} $ and $ \omega_\mathrm{c} $ are renormalized by the drive frequency $ \omega_\mathrm{d} $.
Let $ \left\{\ket{e}, \ket{g} \right\} $ be the orthogonal basis of one qubit, where $ \ket{g} $ is the ground state, and $ \ket{e} $ is the excited state.
We use the computational basis given by $ \left\{\ket{q_1} \otimes \ket{q_2} \otimes \dots \equiv \ket{q_1q_2\dots} \right\} $ with $ q_i \in \left\{e,g\right\} $.

For a two-qubit system, the set of basis states with the corresponding energies of $ H_\mathrm{q} $ with $ \varepsilon_\mathrm{q} = 0 $ is given by:
\begin{equation}
	\begin{aligned}
		\ket{T_{+}} &= \ket{ee}, & E_{T_{+}} &= 2\omega_\mathrm{q},\\
		\ket{S} &= \frac{1}{\sqrt{2}}\left(\ket{eg} - \ket{ge}\right), & E_{S} &= \omega_\mathrm{q},\\
		\ket{T_{0}} &= \frac{1}{\sqrt{2}}\left(\ket{eg} + \ket{ge}\right), & E_{T_{0}} &= \omega_\mathrm{q},\\
		\ket{T_{-}} &= \ket{gg}, & E_{T_{-}} &= 0.
	\end{aligned}
	\label{eq:uncoupled_basis}
\end{equation}
The singlet state $ \ket{S} $ and the triplet state $ \ket{T_0} $ are of special interest as target states, because they correspond to two of the maximally entangled Bell states.
Analogously to the qubit states, we define the photonic basis states by $ \left\{\ket{n_1} \otimes \ket{n_2} \otimes \dots \equiv \ket{n_1n_2\dots} \right\} $, where $ n_i $ is the number of photonic excitations in cavity mode $ i $.
For the joint system of cavities and qubits we use the basis $\{\ket{q_1q_2\dots}\otimes\ket{n_1n_2\dots}\equiv\ket{q_1q_2\dots,n_1,n_2\dots}\}$.
We also use the notation $\ket{S}$ and $\ket{T_0}$ for photonic states in analogy to the qubit states.
In order to formulate the Hamiltonian in matrix representation, we use a reduced state space, in which we divide the basis states of the joint cavity-qubit system into groups distinguished by their total number of excitations.
Furthermore, we divide the groups into subgroups with symmetric or antisymmetric basis states with respect to cavity or qubit permutations, respectively.
This is helpful for using the number of excitations as a cutoff.
To give an example, the matrix representation of the Hamiltonian restricted to the subspace with zero and one excitation reads ($ J_{1,2} = J_{2,1} \equiv J $)
\begin{equation}
	H = 
	\begin{bmatrix}
		0 & \sqrt{2}\varepsilon_\mathrm{q} & 0 & 0 & 0 \\
		\sqrt{2}\varepsilon_\mathrm{q} & \omega_\mathrm{q} - \omega_\mathrm{d} & g & 0 & 0 \\
		0 & g & \omega_\mathrm{c}^{-} - \omega_\mathrm{d} & 0 & 0 \\
		0 & 0 & 0 & \omega_\mathrm{q} - \omega_\mathrm{d} & g \\
		0 & 0 & 0 & g & \omega_\mathrm{c}^{+} - \omega_\mathrm{d}
	\end{bmatrix},
	\label{eq:Tmatrix}
\end{equation}
with $ \omega_\mathrm{c}^\pm \equiv \omega_\mathrm{c} \pm J $ being the symmetric ($ \omega_\mathrm{c}^{-} $) and antisymmetric mode ($ \omega_\mathrm{c}^{+} $) of the coupled-cavity system, respectively.
The rows and columns of the matrix are ordered according to $ \ket{gg,00} $, $ \ket{T_0,00} $, $ \ket{gg,T_0} $, $ \ket{S,00} $, and $ \ket{00,S} $ in the combined qubit and photon state space.
The state $ \ket{gg,00} $ is the cavity-qubit vacuum state (VS).
The other basis states are states with one excitation each.
The coupling of the excitation blocks is controlled by the coherent pump amplitude $ \varepsilon_\mathrm{q} $, while there is no coupling between symmetric and antisymmetric states.
For example, it is not possible to drive the qubits directly from the symmetric vacuum state $ \ket{gg} $ to the antisymmetric state $ \ket{S} $ without generating an additional photonic excitation, so that the total system ends up in the symmetric state $ \ket{S,S} $.
From these symmetry considerations, it becomes clear that the generation of entanglement in the qubit (photonic) subspace must take into consideration the symmetry of the state in that of the photons (qubits). For more details see also Section~\ref{sec:symmetrysteering}.

The eigenvalues of the matrix representation of the Hamiltonian $ H $ in Eq.~\eqref{eq:Tmatrix} for a two cavity-qubit system without an external drive are given by the vacuum state energy $ E_0 = 0 $ and
\begin{equation}
	\begin{aligned}
		E_{1,2} &= \frac{1}{2}\left(\omega_\mathrm{q}+\omega_\mathrm{c}^{-}\right) \pm \frac{1}{2}\tilde{\Omega}^{-},\\
		E_{3,4} &= \frac{1}{2}\left(\omega_\mathrm{q}+\omega_\mathrm{c}^{+}\right) \pm \frac{1}{2}\tilde{\Omega}^{+}.
	\end{aligned}
\end{equation}
The gap between the eigenenergy pairs ($E_1$, $E_2$), and ($E_3$, $E_4$) is given by the generalized Rabi frequency \mbox{$ \tilde{\Omega}^\pm \equiv \sqrt{(\omega_\mathrm{q}-\omega_\mathrm{c}^\pm)^2 + 4g^2} $}.

\begin{figure}
	\centering
	\includegraphics{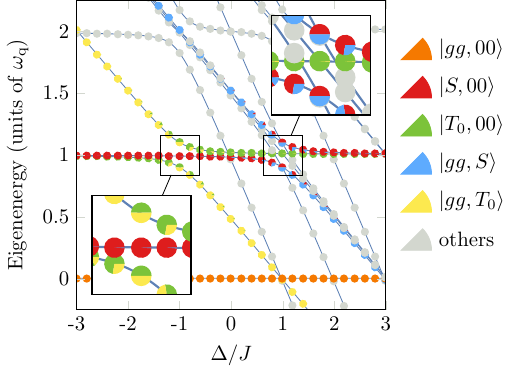}
	\caption{Eigenspectrum representing the eigenenergies and eigenstates of the two cavity-qubit model in dependence of the cavity-qubit detuning $ \Delta $.
	The insets are magnifications of the anticrossing regions, at which the splitting is given by $ \tilde{\Omega}^\pm = 2g $ at $ \Delta = \pm J $.}
	\label{fig:eigenspectrum2_without_pump}
\end{figure}

In order to understand the CCA energy structure and possible excitation pathways of the full system, we introduce a graphical representation of the CCA eigenspectrum as shown in Fig.~\ref{fig:eigenspectrum2_without_pump}.
In the absence of a coherent driving process ($ \varepsilon_\mathrm{q} = 0 $), the eigenenergies of the system's low-excitation states are shown as a function of cavity-qubit detuning.
Each colored disk represents one eigenstate where the size of the colored segments indicates its squared overlap with the basis states introduced in Section~\ref{sec:CCAeigenspectrum}.
The insets show a magnification of the anticrossing regions.
There, the symmetric (green, yellow) and the antisymmetric (red, blue) cavity-qubit states mix in analogy to the well know Jaynes-Cummings splitting, given by $2g$.
However, the anticrossings of symmetric and antisymmetric state pairs occur at different cavity-qubit detunings $\Delta\equiv\omega_\mathrm{q}-\omega_\mathrm{c}$, allows for individually addressability of these subspaces.

For $\Delta=J$, the one-excitation subspace features two eigenstates that possess equal overlap of 0.5 with the states $\ket{S,00}$ (red) and $\ket{gg,S}$ (blue).
These eigenstates are given by $\ket{\psi_{+}}=(\ket{S,00}+\mathrm{e}^{\mathrm{i}\varphi}\ket{gg,S})/\sqrt{2}$, and $\ket{\psi_{-}}=(\ket{S,00}-\mathrm{e}^{\mathrm{i}\varphi}\ket{gg,S})/\sqrt{2}$, where $\varphi$ is a phase that is not contained in the visual representation in Fig.~\ref{fig:eigenspectrum2_without_pump}.
A similar situation is given for $\Delta=-J$, where the one-excitation sector features two eigenstates possessing equal overlap of 0.5 with the states $\ket{T_0,00}$ (green) and $\ket{gg,T_0}$ (yellow).
The vacuum state $ \ket{gg,00} $ (orange) is not coupled to any other state as expected from Eq.~\eqref{eq:Tmatrix} for $\varepsilon_\mathrm{q}=0$.

\section{Entanglement generation in the bipartite system}
\label{sec:egscheme}
\begin{figure}
	\centering
	\includegraphics{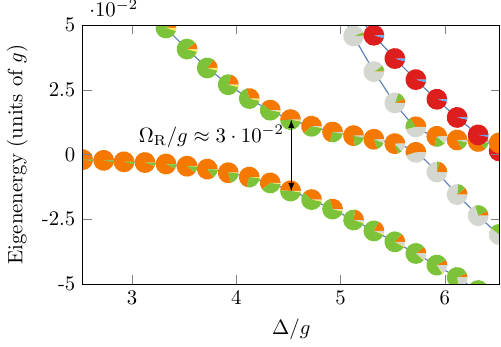}
	\caption{Eigenspectrum for a system with two cavity-qubit systems as a function of their detuning $\Delta$.
	The parameters used here are $ J = g $, $ \omega_\mathrm{q} = 70g $, $ \varepsilon_\mathrm{q} = 0.01g $.
	At $ \Delta = 4.53g $ (i.e.~$\omega_\mathrm{c}=65.47g$) and $\omega_\mathrm{d}=70.17g$ we expect Rabi oscillation of frequency $\Omega_\mathrm{R} \approx 0.03g$ with a maximum amplitude.}
	\label{fig:eigenspectrum2_with_pump}
\end{figure}

\begin{figure}
	\centering
	\includegraphics{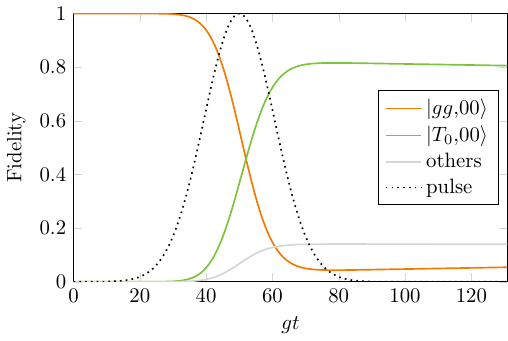}
	\caption{Fidelities of the vacuum state $\ket{gg,00}$, the entangled target state $\ket{T_0,00}$, the sum of the fidelities of all other states, and the normalized Gaussian pulse.
	Initially, the system is in the vacuum state.
	After driving the system with the Gaussian pulse, the fidelity of the target state reaches a value of about 0.84.
	The included dissipation effects have the rates $\kappa=10^{-3}g$, $\gamma=10^{-4}g$, $\gamma_\varphi=10^{-5}g$.}
	\label{fig:pulse}
\end{figure}

In this section, we present a numerical method for exciting target states without further transformations or approximations of the Hamiltonian that was introduced in the previous section.
We begin by identifying possible transitions between the vacuum state and an entangled target state from the eigenenergy spectrum shown in Fig.~\ref{fig:eigenspectrum2_with_pump}.
The idea is to determine transitions between eigenstates that ideally have equal and maximal overlap with the vacuum and the target state.
In this case, coherent excitation can be used to drive this transition without generating additional state-admixtures in the state of the system.
As an example, in Fig.~\ref{fig:eigenspectrum2_with_pump}, such a transition can be identified between the vacuum state $\ket{gg,00}$ and the target state $\ket{T_{0},00}$ at a detuning of $\Delta=4.53g$, where a realistic value for the cavity-qubit coupling is $ g \approx 2\pi \cdot 10^{-1}\,\mathrm{GHz} $ \cite{kimchi-schwartz_stabilizing_2016}. 
This targeted transition can be reached with a $\pi$-pulse, for which the area under the curve describing the pulse function must be the product of the driving strength $\varepsilon_\mathrm{d}$ and half the duration $T_\mathrm{R}=2\pi/\Omega_\mathrm{R}$ of a Rabi cycle.
In the following, we use the fidelity as a measure of how close the system is to the target state.
For two pure states $\ket{\phi}$ and $\ket{\psi}$ it is given by their squared overlap, i.e.~$ F \equiv |\braket{\psi|\phi}|^2 $.
Fig.~\ref{fig:pulse} shows the fidelities of the vacuum state (orange line), which is the initial state, the target state (green line), and the sum of the fidelities of the other states of the system (gray line) for an excitation with a Gaussian $\pi$-pulse (dotted black line) at the identified driving frequency $\omega_\mathrm{d}=70.17g$.
For tailoring the Gaussian pulse, the Rabi frequency $\Omega_\mathrm{R} \approx 0.028g$ that is shown in Fig.~\ref{fig:eigenspectrum2_with_pump} was used.
For the entangled target state $\ket{T_{0},00}$, we obtain a maximum fidelity of $ F_\mathrm{TS} \approx 0.95 $.
The deviation from the optimal value of 1 is caused by the admixture of residual states, here mainly the state $ \ket{gg,T_{0}}$, which can also be inferred from the small contribution (yellow) to the eigenstate in Fig.~\ref{fig:eigenspectrum2_with_pump}.

To shed more light on the mechanism limiting the maximum fidelity of the target state, we approximate the pair of Hamiltonian eigenstates involved in the transition (see Fig.~\ref{fig:eigenspectrum2_with_pump}) in terms of a driven three-level system describing the ground and target state, and a third state representing the combined influence of all other levels in the system. 
Interpreting the symmetric part of the single-excitation subspace of the Hamiltonian in Eq.~\eqref{eq:Tmatrix} in this way an approximate analytic expression of the maximum attainable fidelity for excitation of the W~state can be derived (see Appendix \ref{appendix:maxfidelityswtrafo}), i.e. 
\begin{equation}
	F_\mathrm{max} = \frac{\tilde{\varepsilon}_\mathrm{q}^2}{\left(\frac{\tilde{\Delta}_\mathrm{qd}}{2}\right)^2 + \tilde{\varepsilon}_\mathrm{q}^2}
	\label{eq:approxmaxfidelity}
\end{equation}
with
\begin{equation}
	\begin{aligned}
		\tilde{\varepsilon}_\mathrm{q} &= \sqrt{2}\varepsilon_\mathrm{q} + \frac{\sqrt{2}\varepsilon_\mathrm{q} g^2}{2(2\varepsilon_\mathrm{q}^2 - \Delta_\mathrm{cd}^{-}(\Delta_\mathrm{cd}^{-}-\Delta_\mathrm{qd}))},\\
		\tilde{\Delta}_\mathrm{qd} &= \Delta_\mathrm{qd} + \frac{\Delta_\mathrm{cd}^{-} g^2}{2\varepsilon_\mathrm{q}^2 - \Delta_\mathrm{cd}^{-}(\Delta_\mathrm{cd}^{-}-\Delta_\mathrm{qd})},
	\end{aligned}
\end{equation}
where $ \Delta_\mathrm{qd} \equiv \omega_\mathrm{q} - \omega_\mathrm{d} $ and $ \Delta_\mathrm{cd}^{-} \equiv \omega_\mathrm{c}^{-} - \omega_\mathrm{d} $.
Eq.~\eqref{eq:approxmaxfidelity} shows that the optimal parameters are found when the qubit drive frequency is close to the qubit energy. 
However, the qubit energy is renormalized due to the light-matter interaction with the cavity, which is akin to an AC Stark effect with large detuning.

\begin{figure}
	\centering
	\includegraphics{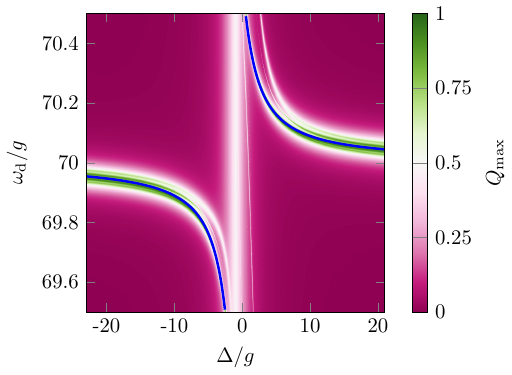}
	\caption{Maximum overlap quality $Q_\mathrm{max}$ for generating the bipartite target state $\ket{T_0,00}$ as a function of cavity-qubit detuning $\Delta$ and the driving frequency $\omega_\mathrm{d}$.
	The blue line represents the analytic result in Eq.~\eqref{eq:analyticapproach}.}
	\label{fig:map2}
\end{figure}

The approach detailed in the previous example can be broken down to the following steps: First, we obtain the eigenstates $ \ket{\psi_i} $ of the system for a given set of parameters.
Next, we calculate the fidelities of these eigenstates with the vacuum and target state, respectively, i.e.~$ F_{\mathrm{VS},i} = |\braket{\psi_i | gg,00}|^2 $ and $ F_{\mathrm{TS},i} = |\braket{\psi_i | T_{0},00}|^2 $.
Since we are ideally looking for an eigenstate with equal contribution of the vacuum and the target state only, the optimum value of their squared overlap with this eigenstate is 0.5.
The absolute differences between the fidelities $ F_{\mathrm{VS},i} $ or $ F_{\mathrm{TS},i} $, and the value of 0.5 are used to introduce the overlap quality $ Q_i $ of the eigenstate $ \ket{\psi_i} $ as
\begin{equation}
	Q_i \equiv 1 - |0.5 - F_{\mathrm{VS},i}| - |0.5 - F_{\mathrm{TS},i}|,
\end{equation}
which is a value between 0 and 1 with 1 being the optimal case.
For the present example, $ Q_i = 1 $ means that the eigenstate is of the form
\begin{equation}
	\ket{\psi_i} = (\ket{gg,00} + \mathrm{e}^{\mathrm{i}\varphi}\ket{T_{0},00})/\sqrt{2},
\end{equation}
where $ \varphi $ is an undetermined phase.
We compute the $ Q_i $ for each eigenstate and select the one which gives the highest value, i.e.
\begin{equation}
	Q_\mathrm{max} \equiv \max_{i}(Q_i).
	\label{eq:qmaxdefinition}
\end{equation}
In Fig.~\ref{fig:map2} we determine $ Q_\mathrm{max} $ by varying the cavity-qubit detuning $ \Delta $ and the driving frequency $ \omega_\mathrm{d} $ while the other parameters are kept fixed.
The resulting parameter map reveals at which pump frequency and detuning between qubit and cavity mode the quality factor $ Q_\mathrm{max} $ becomes optimal.
In addition to the numerical results, it is possible to derive an analytic formula to find the ideal pump frequency to a given cavity-qubit detuning.
The derivation is given in Appendix~\ref{appendix:analyticalapproach} and leads to the following expression:
\begin{equation}
	\omega_\mathrm{d}^{\pm}(\Delta) = \frac{1}{2}\left(2\omega_\mathrm{q} - \Delta - J \pm \sqrt{\left(\Delta + J\right)^2 + 4g^2}\right),
	\label{eq:analyticapproach}
\end{equation}
which is shown as the blue line in Fig.~\ref{fig:map2}.
The white areas in Fig.~\ref{fig:map2} belong to the overlap quality $Q_\mathrm{max}=0.5$ and indicate parameters, for which one eigenstate of the system is composed of either 50\,\% of the vacuum state, the target state, or the mixture of both.
The other 50\,\% of the eigenstate consist of contributions of other basis states with one or more excitations.
Many of the white areas lie off the blue line and belong to transitions to other states.
Optimal parameters for driving the target state can be found in the green regions, where $Q_\mathrm{max}$ is close or equal to 1.
Fig.~\ref{fig:map2} reveals two of the areas, which are correctly predicted by the analytic approach Eq.~\eqref{eq:analyticapproach}.

\begin{figure}
	\centering
	\includegraphics{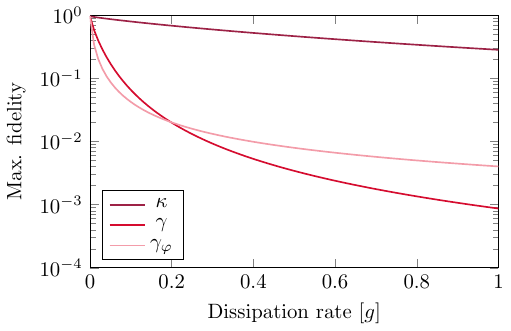}
	\caption{Maximum fidelity with the state $\ket{T_0,00}$ reached after excitation with Gaussian pulse as a function of different dissipation rates $\kappa$, $\gamma$, and $\gamma_\varphi$.
	Qubit dissipation ($\gamma$) and pure dephasing ($\gamma_\varphi$) act directly on the qubit state and thus have a stronger deteriorating effect than cavity decay ($\kappa$).}
	\label{fig:var_dissipation}
\end{figure}

\begin{figure}
	\centering
	\includegraphics{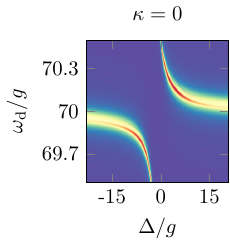}
	\includegraphics{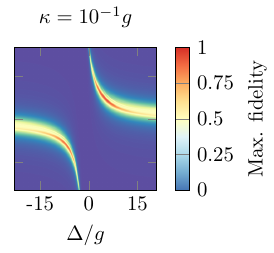}
	\caption{Attainable fidelities for the target state $\ket{T_0,00}$ without (\textbf{left}) and with (\textbf{right}) cavity dissipation ($\kappa=10^{-1}g$) as a function of the cavity-qubit detuning $\Delta$ and the driving frequency $\omega_\mathrm{d}$.
	The red areas indicate parameters for which the fidelity is close to the maximum value of 1.
	The presence of dissipation slightly decreases the size of the red areas and thus the parameter space for entanglement generation.}
	\label{fig:fidelity_map}
\end{figure}

In addition to the coherent dynamics we have addressed so far, irreversible dissipation arises from the non-unitary Liouvillian superoperator in Eq.~\eqref{eq:lindblad}.
The dephasing that is associated with dissipative processes is a well known limitation for entanglement in quantum systems.
We evaluate the impact of pure dephasing, radiative, and cavity losses on the attainable target-state fidelity.
In Fig.~\ref{fig:var_dissipation} the maximum attainable fidelity of the target state $\ket{T_0,00}$ is shown as a function of the dissipation rate for three different dissipation mechanisms.
It can be seen that the physical dissipation mechanism described by the collapse operator $C_n$ significantly impacts the attainable fidelity.
In particular, cavity losses, which can be the predominant loss mechanism in coupled micropillar cavities, have a much weaker impact on the target fidelity than qubit dephasing or radiative losses that directly act on the emitter degrees of freedom.
In Fig.~\ref{fig:fidelity_map}, we augment the parameter map of Fig.~\ref{fig:map2} by directly showing the attainable maximum fidelity for the target state $\ket{T_0,00}$ for the cases without (left) and with small cavity dissipation rate (right).
It can be seen that for values of experimentally attainable dissipation rates, the size of the areas of suitable parameters for entanglement generation are only slightly reduced.

\section{Multipartite entanglement}
\label{sec:mpe}
\begin{figure}
	\centering
	\includegraphics{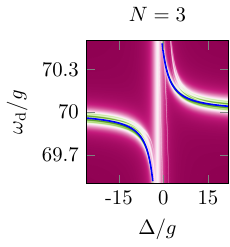}
	\includegraphics{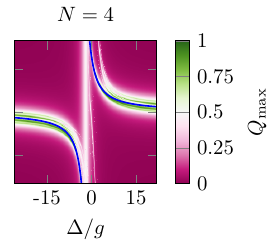}
	\caption{\textbf{Left}: Parameter map as a function of the cavity-qubit detuning $\Delta$ and the driving frequency $\omega_\mathrm{d}$ for generating the tripartite target state $\ket{W_3,000}$.
	We determine $\Delta=3.04g$ (i.e.~$\omega_\mathrm{c}=66.96g$) and $\omega_\mathrm{d}=70.19g$ to be optimal parameters.
	\textbf{Right}: Parameter map for generating the quadripartite target state $\ket{W_4,0000}$.
	We determine $\Delta=2.61g$ (i.e.~$\omega_\mathrm{c}=67.39g$) and $\omega_\mathrm{d}=70.17g$ to be optimal parameters.
	The blue lines represent the analytic result from Eq.~\ref{eq:analyticresultfornccas}.}
	\label{fig:map3map4}
\end{figure}

\begin{figure}
	\centering
	\includegraphics{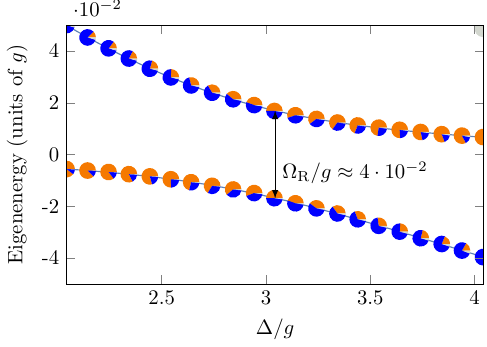}
	\caption{Eigenspectrum of the Hamiltonian with colored charts representing the eigenstates and their squared overlap with the vacuum state $\ket{ggg,000}$, and the target state $\ket{W_3,000}$.
	At $ \Delta = 3.04g $ contributions of the vacuum state (orange) and the target state (blue) are maximal at almost 50\,\%, respectively.
	Gray contributions represent the sum of fidelities of other states.
	The energy gap between the two eigenenergies at the chosen detuning determines the Rabi frequency $ \Omega_\mathrm{R} \approx 0.04g $.}
	\label{fig:eigenspectrum3}
\end{figure}

Quantum networks based on qubits in coupled microcavities are a tangible platform for novel quantum computing and processing concepts \cite{blais_quantum_2020,sun_generation_2016}.
Using the properties enabled by quantum mechanics to the full extent requires control over non-classical correlations and entanglement.
We therefore aim to explore the potential of the approach introduced in the previous section for the generation of multipartite entanglement in systems with more than two coupled cavities.
A relevant multipartite entangled qubit state is the lowest excited Dicke state, or W~state,
\begin{equation}
	\ket{W_N} = \frac{1}{\sqrt{N}}\left(\ket{eg \dots g} + \ket{geg \dots g} + \dots + \ket{g \dots ge}\right).
\end{equation}
W~states are a representative of a class of multipartite entangled states that are particularly robust to loss of particles and global dephasing \cite{briegel_persistent_2001} and represent a vital resource for different quantum communication protocols \cite{buhrman_multiparty_1999}.
Furthermore, for 3 qubit entangled states, the W~states are prototypical for one of two classes of maximally entangled states \cite{dur_three_2000}.
We consider the generation of the W~states for $N=3$ and $N=4$.
For this purpose we use the numerical scheme of the previous section to generate the parameter maps in Fig.~\ref{fig:map3map4} to reveal the excitation space in which the $\ket{W_3,000}$ (left panel) and $\ket{W_4,0000}$ states (right panel) can be driven directly from the corresponding vacuum states $\ket{ggg,000}$ and $\ket{gggg,0000}$.
Again, the blue lines represent the analytical approach given by
\begin{equation}
	\begin{aligned}
		\omega_\mathrm{d}^{\pm}(\Delta) = {} & \frac{1}{2} \bigg( 2\omega_\mathrm{q} - \Delta - (N-1)J \\
		& \pm \sqrt{\left(\Delta + (N-1)J\right)^2 + 4g^2} \bigg)
	\end{aligned}
	\label{eq:analyticresultfornccas}
\end{equation}
for $N=3$ and $N=4$, respectively.
A detailed discussion of this function and its graphical representation is given in Appendix~\ref{appendix:analyticalapproach}.
The parameter maps in Fig.~\ref{fig:map3map4} showing numerical results for the attainable $Q_\mathrm{max}$ for $N=3$ ($N=4$) are shifted by $J$ ($2J$) to the left along the $\Delta$ axis compared to the regions found for $N=2$.
This is confirmed by the analytical approach which shows that the map for a system with $N$ cavities is shifted by $(N-2)J$ compared to the map for $N=2$ cavities.
From the eigenspectrum with the eigenvector representation in Fig.~\ref{fig:eigenspectrum3} we find the Rabi frequency $ \Omega_\mathrm{R} \approx 0.04g $ ($ \Omega_\mathrm{R} \approx 0.04g $) for $N=3$ ($N=4$) cavities in good agreement with the analytical prediction
\begin{equation}
	\Omega_\mathrm{R} = 2 \sqrt{N} \varepsilon_\mathrm{q},
\end{equation}
as detailed in Appendix~\ref{appendix:analyticalapproach}.
By using a Gaussian pulse that is tailored to the correct Rabi frequency, we obtain W~state fidelities of 0.95 for $N=3$ and $N=4$ cavities.
In this section, we have demonstrated the applicability of our generalized
numerical scheme for 3 and 4 cavity-qubit systems, showing that it is possible to specify precise excitation parameters to drive a multipartite system from
the vacuum state directly into a W~state. 
While this approach extends to larger systems in a straightforward way, it quickly becomes numerically expensive. 
We point out, however, that MPE states of 3 and 4 coupled cavity-qubit systems are already quite relevant for some of the emerging quantum technology applications, such as  quantum reservoir computing \cite{fujii_harnessing_2017, mujal_opportunities_2021}.

\section{Symmetry steering via local excitation}
\label{sec:symmetrysteering}
CCAs offer the unique possibility to individually address each lattice site by optical or electrical excitation \cite{sun_generation_2016}.
While, in principle, this offers a plethora of tuning knobs by locally tailoring the temporal and spectral behavior of the pump at each cavity, here we restrict our analysis to the possibility of applying local phase factors to access MPE target states that could not be generated from a homogeneous excitation scheme for the whole array.
A simple example is the generation of the singlet qubit-Bell state $\ket{S,00}$.
As mentioned in Section~\ref{sec:CCAeigenspectrum}, this is not directly possible with the previously used excitation scheme.
However, the generation of an antisymmetric target state from the symmetric vacuum state may be realized by applying a local phase factor to the excitation of a single qubit, or, respectively, a phase difference between the excitation of both qubits.
In case of the state $\ket{S,00}$, the phase difference is $\pi$ and the matrix given in Eq.~\eqref{eq:Tmatrix} becomes
\begin{equation}
	H = 
	\begin{bmatrix}
		0 & 0 & 0 & \sqrt{2}\varepsilon_\mathrm{q} & 0 \\
		0 & \omega_\mathrm{q} - \omega_\mathrm{d} & g & 0 & 0 \\
		0 & g & \omega_\mathrm{c}^{-} - \omega_\mathrm{d} & 0 & 0 \\
		\sqrt{2}\varepsilon_\mathrm{q} & 0 & 0 & \omega_\mathrm{q} - \omega_\mathrm{d} & g \\
		0 & 0 & 0 & g & \omega_\mathrm{c}^{+} - \omega_\mathrm{d}
	\end{bmatrix}.
	\label{eq:Smatrix}
\end{equation}
In comparison to Eq.~\eqref{eq:Tmatrix}, here the vacuum state $\ket{gg,00}$ is coupled to the antisymmetric states including the target state $\ket{S,00}$.
Based on this insight, we can apply the entanglement generation scheme to a more general class of MPE states, which is the subspace spanned by states of the form
\begin{equation}
	\ket{W_N^\mathrm{ph}} = \frac{1}{\sqrt{N}}\left(\mathrm{e}^{\mathrm{i}\phi_1}\ket{eg \dots g} + \dots + \mathrm{e}^{\mathrm{i}\phi_N}\ket{g \dots ge}\right).
\end{equation}
Arbitrary phases $\phi_i$ that are not 0 or $\pi$ lead to additional couplings between the basis states.
This makes the theoretical prediction of where to find areas with suitable parameters for the targeted generation of phased entangled states more sophisticated.
The fact that this is possible at all is a unique feature of the CCA platform with individually addressable sites.
By allowing for different phases of the lasers exciting single cavities, these states can indeed be generated, opening the door for accessing larger parts of the Hilbert space and to go beyond the more conventional entangled states.
We demonstrate this by generating a phased W state for three coupled cavities ($\phi_1=\phi_2=0$, $\phi_3=\pi$), for which a fidelity of about $83\,\%$ can be obtained (see Appendix~\ref{appendix:phasedstates}).

We have restricted the discussion of our scheme to excitations of multipartite entangled qubit states that have at most one excitation.
In principle also multi-excitation states can be targeted but the corresponding Rabi frequencies are reduced on the order in $\Delta/g$ \cite{munoz_emitters_2014} which significantly reduces the attainable fidelities due to the competition with the discussed dephasing mechanisms.

\section{Steady-state entanglement generation via bath engineering}
\label{sec:steadystate}
In this last part, we depart from the time-dependent excitation scheme and consider how \emph{steady-state} entanglement can be obtained from a continuous excitation.
For a steady-state to exist, dissipation must be present to provide a non-unitary contribution to the system dynamics.
The steady-state is then formed by the balance between the continuous coherent drive on the one hand, and dissipation and dephasing on the other hand.
We begin by formulating a Bloch-Redfield approach to show that the steady-state density operator can be expressed in terms of a set of rate equations for its diagonal elements when expressed in the Hamiltonian's eigenbasis.
Consequently, an entangled steady-state can only form if the Hamiltonian possesses at least one entangled eigenstate, and if the transition rates between the eigenstates favor the occupation of one of these as the stationary state.

Up to now, we have considered an excitation process that acts on the emitters, which is an appropriate description e.g.~for optical excitation of micropillar laser arrays \cite{heuser_developing_2020}.
To enable a direct comparison to previous work later in this section, we now consider pumping of cavity photons instead.
We stress that both excitation methods are unitarily equivalent up to a scaling factor, as we show in Appendix~\ref{appendix:pump_equivalence}.
In this scenario, the qubit part of the Hamiltonian reads
\begin{equation}
	H_\mathrm{q} = \sum_{i=1}^{2} \omega_\mathrm{q} \sigma_i^{+} \sigma_i^{-},
\end{equation}
whereas the cavity part is augmented by the direct cavity pump
\begin{equation}
	\begin{aligned}
		H_\mathrm{c} = {} & \sum_{i=1}^2 \left[ \omega_\mathrm{c} a_i^\dagger a_i + 2\varepsilon_\mathrm{c}\cos\left(\omega_\mathrm{d}t\right)(a_i^\dagger + a_i) \right] \\
		& - J(a_1^\dagger a_2 + a_2^\dagger a_1),
	\end{aligned}
\end{equation}
where $\varepsilon_\mathrm{c} = \sqrt{(J-\omega_\mathrm{c} +\omega_\mathrm{d})^2 + (\kappa/2)^2}\varepsilon_\mathrm{q}/g$ is the cavity driving amplitude.
The cavity-qubit interaction Hamiltonian $H_\mathrm{c,q}$ remains unchanged.
We consider a parameter regime, in which the detuning is large, and the dissipation is small compared to the light-matter coupling.
Furthermore, the dissipative parameters obey $\kappa \gg \gamma \gg \gamma_\varphi$.
\begin{figure*}
	\centering
	\includegraphics[width=\textwidth]{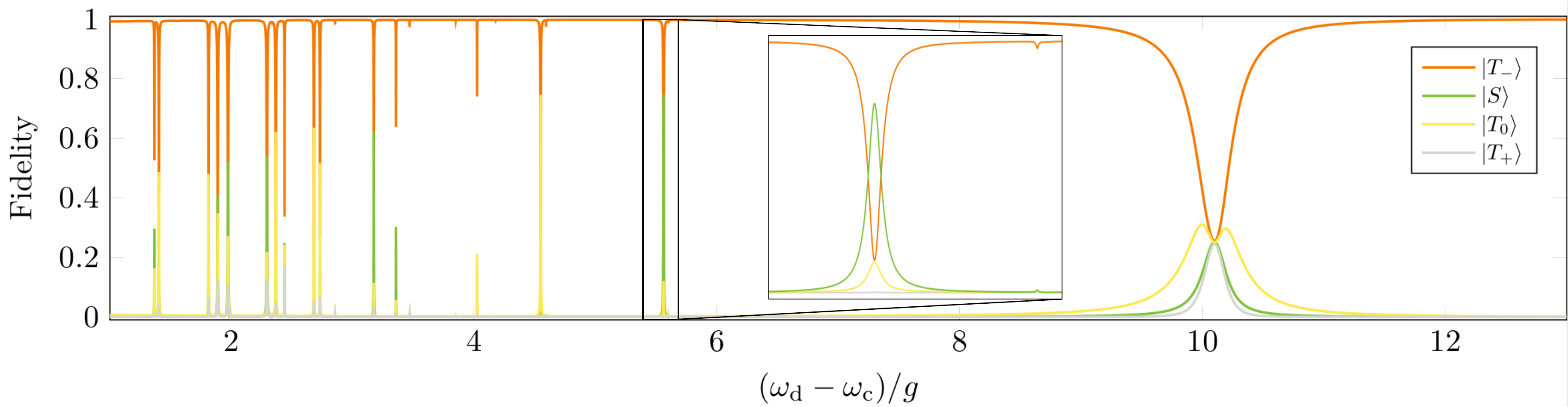}
	\caption{Fidelity of the steady-state for two qubits with the basis states in Eq.~\eqref{eq:uncoupled_basis} as a function of the driving frequency $\omega_\mathrm{d}$.
	The parameters we used are similar to those used \cite{aron_steady-state_2014} and read $\omega_\mathrm{c} = 60g$, $\omega_\mathrm{q} = 70g$, $J=g$, $\varepsilon_\mathrm{c} = g$, $\kappa = 10^{-3}g$, $\gamma = 10^{-4}g$, $\gamma_\varphi = 10^{-5}g$.
	Fidelities represent the direct overlap of the basis states with the reduced density matrix $\rho_\mathrm{q}$ of the qubits.}
	\label{fig:brute_force}
\end{figure*}
For small enough systems, it is feasible to directly compute the steady-state $\rho_\mathrm{ss}$ for different sets of parameters and choose those that are entangled.

The requirement for a steady-state is that the time derivative
\begin{equation}
	\frac{\mathrm{d}}{\mathrm{d}t} \rho_\mathrm{ss} = \mathcal{L} \rho_\mathrm{ss} = 0 
\end{equation}
vanishes.
The Liouvillian superoperator $\mathcal{L}$ describes the time evolution as given by the Lindblad master equation.
In Fig.~\ref{fig:brute_force}, we varied the drive frequency $\omega_\mathrm{d}$ and determined the steady-state for different sets of parameters by finding the eigenvector of $\mathcal{L}$ with eigenvalue 0.
At $\omega_\mathrm{d} - \omega_\mathrm{c} \approx 10g$, the drive is in resonance with the qubit energies.
As a result, the steady-state is the maximally mixed state, where the coherence of the qubit state is lowest and the purity $\mathrm{tr}\, \rho_\mathrm{q}^2 = 0.25$ takes on its minimal value.
At lower driving frequencies we find a large number of resonances at which the stationary state diverges from the qubit ground state and exhibits high fidelity of either the $\ket{T_0}$ or $\ket{S}$ state (see inset for magnification). Being interested in generating steady-state entanglement, we take a closer look at the underlying processes that determine the position of the observed resonances, which show that steady-state entanglement can be found for a wide range of system parameters.

In the literature, obtaining entanglement in the steady-state has been discussed in the context of quantum bath engineering \cite{lee_environment_2019, ma_coupling-modulationmediated_2021} and for coupled cavity arrays specifically \cite{aron_steady-state_2014}.
We provide insight in how it is related to the method we have described above and highlight the underlying principles as well as differences to other approaches.
Although it is possible to find the steady-state by exact diagonalization of the Liouvillian, this approach does not provide insight into the mechanisms that distinguish the different observed resonances.

With the aim to find a more general description for the generation of steady-state entanglement in open quantum systems, we now consider the general form of a system-bath interaction Hamiltonian
\begin{equation}
	H_\mathrm{int} = \sum_\alpha A_\alpha \otimes B_\alpha,
	\label{eq:h_int_general}
\end{equation}
where $A_\alpha$ and $B_\alpha$ are operators solely acting on the system or bath, respectively.
In the Born-Markov approximation, the time evolution of the system is described by the Bloch-Redfield master equation
\begin{equation}
	\frac{\mathrm{d}}{\mathrm{d}t} \rho = - \mathrm{i} [H, \rho] + \sum_{\alpha, \omega} \gamma_\alpha (\omega) \mathcal{D}[A_\alpha(\omega)]\rho
\end{equation}
with Lindblad-type dissipators $\mathcal{D}$.
The jump operators $A_\alpha(\omega_{ij}) = \ket{j}\bra{j}A_\alpha \ket{i}\bra{i}$ are the eigenoperators for each transition frequency $\omega_{ij} = E_i - E_j$ and describe transitions between the energy eigenstates $\ket{i}$ and $\ket{j}$, while $\gamma_\alpha(\omega) = \int_{-\infty}^\infty \mathrm{d}t\, \mathrm{e}^{\mathrm{i}\omega t} \braket{B_\alpha^\dagger(t) B_\alpha(0) }$ is the spectral function of the bath operators \cite{breuer_theory_2002}.

Assuming a non-degenerate Hamiltonian and denoting the matrix elements by $\bra{j}A_\alpha \ket{i} \equiv A^\alpha_{ji}$, we can find a set of linear equations for the matrix elements of the density operator in the eigenbasis of the Hamiltonian:
\begin{align}
	\dot{\rho}_{kk} &= \sum_{\alpha,i} \left[ \gamma_\alpha(\omega_{ik}) |A^\alpha_{ik}|^2 \rho_{ii} -\gamma_\alpha(\omega_{ki})|A^\alpha_{ki}|^2 \rho_{kk} \right],
	\label{eq:diagonals}\\
	\dot{\rho}_{kl} &= - \left[ \sum_\alpha \lambda^\alpha_{kl} + \mathrm{i} \omega_{kl} \right] \rho_{kl} \quad \mathrm{for} \quad k\neq l.
	\label{eq:non_diagonals}
\end{align}
The complete derivation and the exact form of the constant $\lambda^\alpha_{kl}$ can be found in Appendix \ref{appendix:rates}.
Eq.~\eqref{eq:non_diagonals} shows that the equations for off-diagonal elements decouple from each other, and that they decay exponentially. This leads to a steady-state density matrix that is diagonal in the eigenbasis of the Hamiltonian.
From this follows the important result that in order to achieve steady-state entanglement, we need the Hamiltonian to have entangled eigenstates, since a classical mixture of non-entangled states cannot produce entanglement.
The other relevant factor for entanglement of the steady-state is the occupation $\rho_{kk}$ of an entangled state $\ket{k}$, which can be calculated from Eq.~\eqref{eq:diagonals}.
These coupled, linear differential equations can be interpreted as a set of rate equations, with the rate from state $\ket{k}$ to $\ket{l}$ being $\Gamma_{k \rightarrow l} = \sum_\alpha \gamma(\omega_{kl})|A_{kl}^\alpha|^2$.
In this context, one may interpret the influence of the driven cavities on the same footing as the dissipative rates, which has previously led to the formulation of the term ``quantum bath engineering'' to tailor the rates in Eq.~\eqref{eq:diagonals} \cite{aron_steady-state_2014}.

In the following, we discuss how to connect the approach in Ref.~\cite{aron_steady-state_2014} to the general formalism we have detailed above.
We begin by introducing a set of approximations that allow us to simplify our system in order to describe it by using a Bloch-Redfield master equation. This will enable us to understand the mechanism that leads to steady-state entanglement.
We first switch to the rotating frame and perform the rotating wave approximation to eliminate the explicit time dependence.
Furthermore, we apply the Schrieffer-Wolff transformation to separate qubits and cavities up to second order in $g/\Delta$ and use the mean-field approximation to eliminate all non-linearities in the light matter coupling \cite{aron_steady-state_2014}.
A detailed discussion can be found in Appendix~\ref{appendix:approximations}.
The resulting qubit Hamiltonian has the form
\begin{equation}
	\tilde{H}_\mathrm{q} = \frac{1}{2} \sum_{i=1}^2 \left[ \Omega_\mathrm{R} \sigma^x_i + \tilde{\Delta}\sigma^z_i \right] - \frac{1}{2}J \left( \frac{g}{\Delta} \right)^2 (\sigma_1^+ \sigma_2^- + \sigma_2^+ \sigma_1^-)
\end{equation}
with the Rabi frequency $\Omega_\mathrm{R} = 2\varepsilon_\mathrm{c}g/(\Delta+J)$ and the renormalized qubit energies $\tilde{\Delta}$ given by Eq.~\eqref{eq:renormqenergies}.
Importantly, $H_\mathrm{q}$ now includes an \emph{effective qubit-qubit interaction},
leading to eigenstates and energies that are different from those of the pure (untransformed) qubit Hamiltonian:
\begin{equation}
	\begin{aligned}
		\ket{\tilde{T}_{+}} &\approx \ket{T_{+}} + \frac{\Omega_\mathrm{R}}{\sqrt{2}\tilde{\Delta}}\ket{T_0}, &E_{\tilde{T}_{+}} &\approx \tilde{\Delta} + \frac{\Omega_\mathrm{R}^2}{2\tilde{\Delta}},
		\\
		\ket{\tilde{S}} &= \ket{S}, &E_{\tilde{S}} &=J\left( \frac{g}{\Delta} \right)^2,
		\\
		\ket{\tilde{T}_0} &\approx \ket{T_0} + \frac{\Omega_\mathrm{R}}{\sqrt{2}\tilde{\Delta}}(\ket{T_{-}} - \ket{T_{+}}), &E_{\tilde{T}_0} &\approx - J\left( \frac{g}{\Delta} \right)^2,
		\\
		\ket{\tilde{T}_{+}} &\approx \ket{T_{-}} - \frac{\Omega_\mathrm{R}}{\sqrt{2}\tilde{\Delta}}\ket{T_0}, &E_{\tilde{T}_{+}} &\approx - \tilde{\Delta} - \frac{\Omega_\mathrm{R}^2}{2\tilde{\Delta}}.
	\end{aligned}
\end{equation}
In the parameter regime	we use (see caption to Fig. \ref{fig:brute_force}), $\Omega_\mathrm{R}/(\sqrt{2}\tilde{\Delta}) \ll 1$ and therefore the eigenstate $\ket{\tilde{T}_0}$ is close to the maximally entangled state $\ket{T_0}$.
More importantly, the previously degenerate states $\ket{T_0}$ and $\ket{S}$ are now exhibiting an energy splitting given by $2J(g/\Delta)^2$, making them distinguishable in the system dynamics and individually addressable

The goal is now to reach a steady-state that has a high fidelity of either the $\ket{\tilde{T}_0}$ state (and thus of $\ket{T_0}$), or of the $\ket{S}$ state, both of which are entangled.
The advantage of this approach is that the Bloch-Redfield equation provides rates between these states without any coherent mixing between them.
These rates follow from the residual qubit-photon interaction, which is not included in the effective qubit-qubit interaction and is described by the Hamiltonian
\begin{equation}
	\tilde{H}_\mathrm{c,q} = (c D^\dag + c^\ast D)(\sigma^z_1 + \sigma^z_2) + (c d^\dag + c^\ast d)(\sigma^z_1 - \sigma^z_2).
	\label{eq:h_cq_tilde}
\end{equation}
Here, the operators $D^{(\dag)}$ and $d^{(\dag)}$ are annihilation (creation) operators for symmetric and antisymmetric photon fluctuation modes, respectively, with the mode frequencies $\omega_\mathrm{c}^\mp = \omega_\mathrm{c} \mp J$.
They obey the bosonic commutator relations, in particular $[D,D^\dagger]=1$ and $[d,D^\dagger]=0$.
The parameter $c$ describes the strength of the residual qubit-photon interaction and is given in Appendix \ref{appendix:approximations}.
We identify the operators in Eq.~\eqref{eq:h_cq_tilde} with the operators in Eq.~\eqref{eq:h_int_general} as $A_1 = \sigma^z_1 + \sigma^z_2$ and $A_2 = \sigma^z_1 - \sigma^z_2$, $B_1 = c D^\dag + c^\ast D $ and $B_2 = c d^\dag + c^\ast d$.

\begin{figure}
	\centering
	\includegraphics{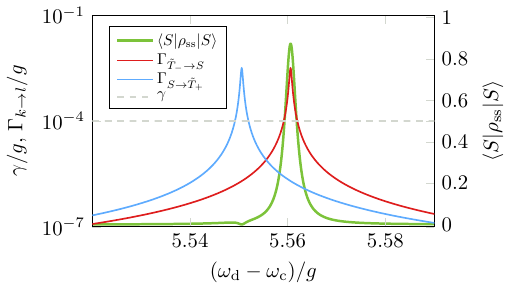}
	\caption{Transition rates $\Gamma_{\tilde{T}_{-} \rightarrow S}$ (red) and $\Gamma_{S \rightarrow \tilde{T}_{+} }$ (blue) around the resonance $\omega_{\tilde{T}_{-}S} = \omega_\mathrm{c}^+ - \omega_\mathrm{d}$ and the resulting steady-state singlet fidelity $\bra{S}\rho_\mathrm{ss}\ket{S}$ (green) as a function of the driving frequency $\omega_\mathrm{d}$.}
	\label{fig:rates}
\end{figure}

In the following, we focus on the influence of the spectral function on the transition rates.
An in-depth analysis including the matrix elements can be found in Ref. \cite{aron_steady-state_2014}.
The spectral functions are proportional to the density of states of the photonic modes
\begin{equation}
	\gamma_{d/D}(\omega) = 2\pi |c|^2 \rho_\pm(\omega),
\end{equation}
with 
\begin{equation}
	\rho_\pm (\omega) = \frac{\kappa / (2\pi)}{(\omega - (\omega_\mathrm{c}^\pm-\omega_\mathrm{d}))^2 + (\kappa/2)^2}.
\end{equation}
This means that the transition rate between two states $\ket{k}$ and $\ket{l}$ is at its maximum if the transition frequency is resonant with the mode frequency, i.e.~$\omega_{kl} = \omega_\mathrm{c}^\pm-\omega_\mathrm{d}$.\\
We take the steady-state generation of the $\ket{S}$ state as an example.
To get a strong transition from the ground state $\ket{\tilde{T}_{-}}$ to $\ket{S}$, the resonance condition $\omega_{\tilde{T}_{-}S} = \omega_\mathrm{c}^+ - \omega_\mathrm{d}$ has to be fulfilled.
In Fig.~\ref{fig:rates} we can see how the singlet fidelity increases up to a maximum of $\bra{S}\rho_\mathrm{ss}\ket{S} = 0.82$ as the rate $\Gamma_{\tilde{T}_{-} \rightarrow S}$ rises above the qubit decay rate $\gamma$ indicated by the gray dashed line.
Next to it, we see the rate $\Gamma_{S \rightarrow \tilde{T}_{+} }$ whose maximum is shifted by $J(g/\Delta)^2$ from the maximum of $\Gamma_{\tilde{T}_{-} \rightarrow S}$.
If $\omega_\mathrm{d}$ is chosen such that the transition $\ket{S} \rightarrow \ket{\tilde{T}_{+}} $ is resonant with the cavity mode, the realization probability of the $\ket{S}$ state reduces rapidly, leading to a dip in steady-state fidelity (green curve in Fig~\ref{fig:rates}).
We point out the importance of the second-order energy corrections following from the Schrieffer-Wolff transformation.
Without it, the transitions $\omega_{\tilde{T}_{-} S}$ and $\omega_{S \tilde{T}_{+}}$ would be degenerate, and we would not be able to distinguish them in this analytic approach based on the Bloch-Redfield equation.

\begin{figure}
	\centering
	\includegraphics{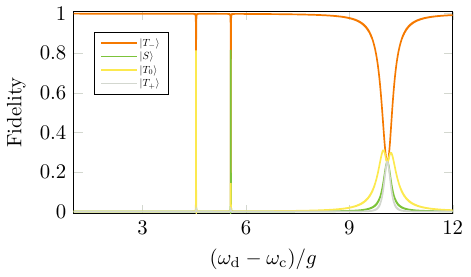}
	\caption{Fidelities of singlet and triplet states in the steady-state for two qubits as a function of the driving frequency $\omega_\mathrm{d}$ for the system after applying the Schrieffer-Wolff transformation.
	The parameters used are the same as in Fig.~\ref{fig:brute_force}}
	\label{fig:brute_force_sw}
\end{figure}

Finally, we investigate the impact of the approximations that were involved in introducing the effective quantum bath engineering picture.
By performing the Schrieffer-Wolff transformation, the remaining light-matter interaction has a reduced strength that is only of second order in $g/\Delta$.
This allows one to make use of the Bloch-Redfield rate formalism to reduce the problem of steady-state entanglement generation to the optimization of transition rates between eigenstates.
These approximations naturally reduce the number of possible entangled eigenstates contained in the full Hamiltonian by separating the qubits from the photonic degrees of freedom.
To assess the impact of the involved approximations, we compare the results involving the Schrieffer-Wolff approximation with those obtained from the full numerical solution.
To do so, we investigate the Schrieffer-Wolff Hamiltonian for entangled steady-states in the same fashion that led to Fig.~\ref{fig:brute_force} with the results shown in Fig.~\ref{fig:brute_force_sw}.
Surprisingly, we see that most peaks that are visible in Fig.~\ref{fig:brute_force} are missing.
Only two peaks remain, the right one is the one explained beforehand as means to generate the $\ket{S}$ state (compare Fig.~\ref{fig:rates}) and the left corresponds to an excitation frequency that results in a high fidelity of $\ket{\tilde{T}_0}$ that can be explained along the same lines, using the resonance condition $\omega_{\tilde{T}_{-}\tilde{T}_0} = \omega_\mathrm{c}^{-} - \omega_\mathrm{d}$.

By comparing Fig.~\ref{fig:brute_force} and \ref{fig:brute_force_sw}, we conclude that the additional peaks that are present in Fig.~\ref{fig:brute_force} must correspond to higher order resonances in $g/\Delta$ that possess a lower fidelity of the entangled target states.
Even though the Schrieffer-Wolff transformation neglects these higher order processes, the resonances that lead to the highest fidelities of the entangled states $\ket{S}$ and $\ket{T_0}$ are contained already in second order with the additional benefit of providing an understanding on how the steady-state is formed.
However, one should keep in mind that the actual physical system contains a richer physics that is lost in the approximation procedure, and steady-state entanglement can in principal be achieved in a much wider parameter range.

\section{Conclusion}
\label{sec:conclusion}
Coupled cavity arrays have been recognized as a promising platform for hosting entangled states.
Tunability of the light-matter interaction and the inter-cavity couplings enable a large degree of control over system properties.
At the same time, each cavity can be individually addressed to realize excitation schemes with locally tuned phases.
In this context, we analyze the direct driving of multipartite entangled target states from the vacuum state by means of detuned optical excitation pulses as well as continuous optical excitation into an entangled steady-state.
For homogeneous few-cavity-qubit systems, we lay out paths for the generation of Bell and W~states, as well as phased W~states by combining analytical and full numerical methods.
From the insight we obtain, a generalized numerical scheme is derived to identify suitable driving parameters for different classes of entangled target states.
Moreover, we connect our approach to the concept of quantum bath engineering, which is an intriguing concept especially to drive complex multipartite systems adiabatically into entangled eigenstates.
While analytical approaches to quantum bath engineering provide a deep understanding of the underlying mechanism, namely by introducing an effective qubit-qubit coupling that leads to entangled eigenstates of the Hamiltonian, our analysis also reveals that the involved approximations sacrifice higher-order processes that lead to additional means for entanglement generation.

This work constitutes a first step towards more complex network topologies based on CCAs with individually tailored inter-cavity couplings.
The developed scheme allows for inducing quantum correlations related to entanglement in a controlled way, which is a key resource for applications in photonic quantum technologies, such as quantum reservoir computing.

\begin{acknowledgments}
	We are thankful to Paul Gartner for useful discussions.
	This work has been supported by the Deutsche Forschungsgemeinschaft (DFG) via the graduate school 2247 ``\textit{Quantum Mechanical Materials Modelling}'', and project Gi1121/6-1 ``\textit{Photonic Quantum Reservoir Computing}''.
	We further acknowledge funding from the Bundesministerium f\"{u}r Bildung und Forschung (BMBF) via the QuanterERA II European Union's Horizon 2020 research and innovation programme under the EQUAISE project, Grant Agreement No. 101017733.
	F.~Lohof is grateful for funding from the Central Research Developing Funds of the University of Bremen.
\end{acknowledgments}

\appendix

\section{Maximum fidelity estimation}
\label{appendix:maxfidelityswtrafo}
\begin{figure}
	\centering
	\includegraphics{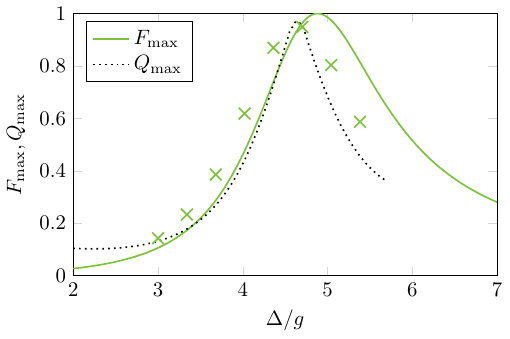}
	\caption{Maximum attainable fidelity $ F_\mathrm{max} $ for the target state 
	$ \ket{T_0,00} $, as a function of detuning. 
	The approximation (green line) and the numerical results (crosses) show the same behavior. The overlap quality $ Q_\mathrm{max} $ from Eq.~\eqref{eq:qmaxdefinition} is shown in black.}
	\label{fig:swtfidelity}
\end{figure}

We derive an estimate for the maximum fidelity of the entangled target-state generation using the optical driving scheme introduced in Sec.~\ref{sec:egscheme}. 
Restricting the Hamiltonian in Eq.~\eqref{eq:Tmatrix} to the single-excitation subspace of all symmetric states (under permutation of the cavity-qubit subsystems), the Hamiltonian can be written in terms of an effective two-level system weakly coupled to a perturbing third level, i.e.~$H=H_0+V$, where 
\begin{equation}
    H_0 = \left(
    \begin{matrix}
        0 & \sqrt{2}\varepsilon_\mathrm{q} & 0 \\
        \sqrt{2}\varepsilon_\mathrm{q} & \Delta_\mathrm{qd} & 0 \\
        0 & 0 & \Delta_\mathrm{cd}^{-}
    \end{matrix}
    \right), \quad 
    V = \left(
    \begin{matrix}
        0 & 0 & 0 \\
        0 & 0 & g \\
        0 & g & 0
    \end{matrix}
    \right)
\end{equation} 
are the uncoupled ($H_0$) and the coupling Hamiltonian ($V$) connecting the subspace of ground- and target-state to the perturbing third level. 
We apply a Schrieffer-Wolff transformation $H'=e^{S}He^{-S}$ to this effective Hamiltonian, using 
\begin{equation}
    S = \left(
    \begin{matrix}
        0 & 0 & S_{13} \\
        0 & 0 & S_{23} \\
        -S_{13} & -S_{23} & 0
    \end{matrix}
    \right)
\end{equation}
with 
\begin{equation}
	\begin{aligned}
		S_{13} &= \frac{\sqrt{2}\varepsilon_\mathrm{q} g}{2\varepsilon_\mathrm{q}^2 - \Delta_\mathrm{cd}^{-}(\Delta_\mathrm{cd}^{-} - \Delta_\mathrm{q})},\\
		S_{23} &= \frac{\Delta_\mathrm{cd}^{-} g}{2\varepsilon_\mathrm{q}^2 - \Delta_\mathrm{cd}^{-}(\Delta_\mathrm{cd}^{-} - \Delta_\mathrm{q})},
	\end{aligned}
\end{equation} 
obtained by solving $[H_0,S]=V$. This allows an expansion of the transformed Hamiltonian such that the first two levels become decoupled from the third one up to first order in $V$, i.e.~$H' = H_0 + \frac{1}{2}[S,V] + \mathcal{O}(V^3)$. 
This results in an effective two-level system with the Hamiltonian 
\begin{equation}
    \tilde{H} = \left(
        \begin{matrix}
            0 & \tilde{\varepsilon}_\mathrm{q} \\
            \tilde{\varepsilon}_\mathrm{q} & \tilde{\Delta}_\mathrm{qd}
        \end{matrix}
        \right),
\end{equation}
in which the detuning and coupling strength are renormalized by the presence of the third level according to 
\begin{equation}
	\begin{aligned}
		\tilde{\varepsilon}_\mathrm{q} &= \sqrt{2}\varepsilon_\mathrm{q} + \frac{\sqrt{2}\varepsilon_\mathrm{q} g^2}{2(2\varepsilon_\mathrm{q}^2 - \tilde{\Delta}_\mathrm{cd}^{-}(\tilde{\Delta}_\mathrm{cd}^{-}-\Delta_\mathrm{qd}))},\\
		\tilde{\Delta}_\mathrm{qd} &= \Delta_\mathrm{qd} + \frac{\Delta_\mathrm{cd}^{-} g^2}{2\varepsilon_\mathrm{q}^2 - \Delta_\mathrm{cd}^{-}(\Delta_\mathrm{cd}^{-}-\Delta_\mathrm{qd})}.
	\end{aligned}
\end{equation}
A calculation of the time evolution of a two-level system subject to the Hamiltonian $\tilde{H}$ that is initially in the vacuum state finally yields the expression for the maximum fidelity of the target state 
\begin{equation}
    F_\mathrm{max} = \frac{\tilde{\varepsilon}_\mathrm{q}^2}{\left(\frac{\tilde{\Delta}_\mathrm{qd}}{2}\right)^2 + \tilde{\varepsilon}_\mathrm{q}^2},
	\label{eq:swtfmax}
\end{equation}
that is given in the main text.
In Fig.~\ref{fig:swtfidelity} we evaluate the maximum fidelity $ F_\mathrm{max} $ for the two-qubit system as a function of cavity-qubit detuning around the optimal parameters for excitation of the target state $ \ket{T_0,00} $. 
The approximation in Eq.~\eqref{eq:swtfmax} (green line) follows the trend of the numerical results (crosses). 
For comparison, the overlap quality $ Q_\mathrm{max} $ defined in Eq.~\eqref{eq:qmaxdefinition} is shown in black. 
The approximation is less accurate for larger detunings as the influence of non-target states increases as seen in Fig.~\ref{fig:eigenspectrum2_with_pump}.

\section{Analytic approximations for optimal driving frequencies}
\label{appendix:analyticalapproach}
To facilitate an analytic approach, we limit the state space of the two-cavity system to the vacuum state and the two symmetric states with one excitation, the target state and the entangled state $ \ket{gg,T_0} $.
Assume the vector $ \ket{\psi} = (C_{gg,00}, C_{T_0,00}, C_{gg,T_0}) $ being the quantum state of the system, where the $C_i$ are time-dependent coefficients corresponding to the three basis states we are considering.
Then, the Schr\"{o}dinger equation can be written in the form
\begin{equation}
	\begin{aligned}
		\mathrm{i}\dot{C}_{gg,00} &= \sqrt{2}\varepsilon_\mathrm{q}C_{T_0,00}, \\
		\mathrm{i}\dot{C}_{T_0,00} &= \sqrt{2}\varepsilon_\mathrm{q}C_{gg,00} + \Delta_\mathrm{qd}C_{T_0,00} + gC_{gg,T_0}, \\
		\mathrm{i}\dot{C}_{gg,T_0} &= gC_{T_0,00} + \Delta_\mathrm{cd}^{-}C_{gg,T_0},
	\end{aligned}
\end{equation}
with $\Delta_\mathrm{qd}\equiv\omega_\mathrm{q}-\omega_\mathrm{d}$ and $\Delta_\mathrm{cd}^{-}\equiv\omega_\mathrm{c}-\omega_\mathrm{d}-J$.
We assume that $|\Delta_\mathrm{cd}^{-}| \gg |\Delta_\mathrm{qd}|, |\varepsilon_\mathrm{q}|, |g|$ to ensure that the system essentially remains in the subsystem of the vacuum and the target state.
In this case $\dot{C}_{00,T_0}\approx 0$ and we eliminate this coefficient from the first two equations, which then form the desired effective Hamiltonian for the vacuum state and the target state:
\begin{equation}
	H_\mathrm{eff} = 
	\begin{bmatrix}
		0 & \frac{1}{2}\Omega_\mathrm{R}^\mathrm{eff} \\
		\frac{1}{2}\Omega_\mathrm{R}^\mathrm{eff} & \Delta_\mathrm{qd} - \frac{g^2}{\Delta_\mathrm{cd}^{-}}
	\end{bmatrix}.
\end{equation}
The maximum amplitude of the Rabi oscillation with the effective Rabi frequency $ \Omega_\mathrm{R}^\mathrm{eff} = 2\sqrt{2}\varepsilon_\mathrm{q} $ is achieved at resonance, i.e.~for $ \Delta_\mathrm{qd} - g^2/\Delta_\mathrm{cd}^{-} = 0 $.
That condition implies the analytic expression for the pumping frequency that is used in the main text
\begin{equation}
	\omega_\mathrm{d}^{\pm} = \frac{1}{2}\left(2\omega_\mathrm{q} - \Delta - J \pm \sqrt{\left(\Delta + J\right)^2 + 4g^2}\right).
	\label{eq:adieliresulttwo}
\end{equation}
Note that $ \omega_\mathrm{d}^{\pm} $ does not depend on the pumping strength $ \varepsilon_\mathrm{d} $.
However, the Rabi frequency depends on it, and with it the pump pulse that drives the system into the target state.

\begin{figure}
	\centering
	\includegraphics{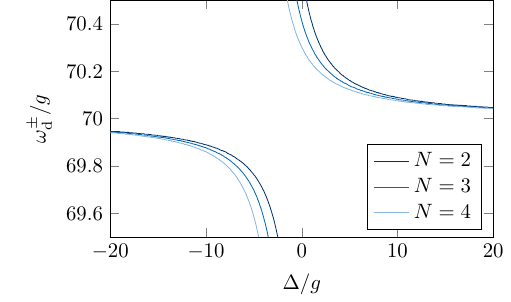}
	\caption{Plot of the driving frequency $\omega_\mathrm{d}^\pm$ given by Eq.~\eqref{eq:adieliresultn} in dependence of the detuning $\Delta$ for $ N = 2, 3, 4 $, $ \omega_\mathrm{q} = 70g $, and $ J = g $.}
	\label{fig:mapoverview}
\end{figure}

The derivation above can be extended to $N$ coupled cavity-qubit systems.
To do so, one has to consider the subsystem of the vacuum state $\ket{g\dots,0\dots}$, the targeted qubit W~state $\ket{W_N,0\dots}$, and the analogous cavity state $\ket{g\dots,W_N}$.
In doing so, we obtain
\begin{equation}\label{eq:omegaeff}
	\Omega_\mathrm{R}^\mathrm{eff} = 2 \sqrt{N} \varepsilon_\mathrm{q}
\end{equation}
and
\begin{equation}
	\begin{aligned}
		\omega_\mathrm{d}^{\pm} = {} & \frac{1}{2} \bigg( 2\omega_\mathrm{q} - \Delta - (N-1)J \\
		& \pm \sqrt{\left(\Delta + (N-1)J\right)^2 + 4g^2} \bigg),
	\end{aligned}
	\label{eq:adieliresultn}
\end{equation}
which are used in Section \ref{sec:mpe} in the main text.
Figure~\ref{fig:mapoverview} shows the plot of Eq.~\eqref{eq:adieliresultn} exemplarily for $ N = 2, 3, 4 $.
The two asymptotes of $\omega_\mathrm{d}^{\pm}$ are given by $\omega_\mathrm{d}^{+} = \omega_\mathrm{q}$ and \mbox{$\omega_\mathrm{d}^{-} = \omega_\mathrm{q}-\Delta-\left(N-1\right)J$}.

\section{Parameter maps for phased W states}
\label{appendix:phasedstates}
\begin{figure}
	\centering
	\includegraphics{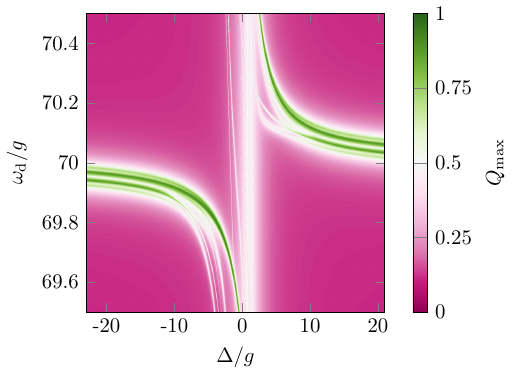}
	\caption{Parameter map as a function of the cavity-qubit detuning $\Delta$ and the driving frequency $\omega_\mathrm{d}$ for generating the phased W state $\ket{W_3^\mathrm{ph},000}$ given by Eq.~\ref{eq:phasedwstate}. We determine as optimal parameters $\Delta=-4.71g$ and $\omega_\mathrm{d}=69.83g$.}
	\label{fig:map3phased}
\end{figure}

In Section \ref{sec:symmetrysteering}, we explore the possibility of generating phased W states using local coherent excitation with individually tunable phases at each lattice site.
In Fig.~\ref{fig:map3phased}, parameter maps are shown for the state 
\begin{equation}
	\ket{W_3^\mathrm{ph},000} = \frac{1}{\sqrt{3}} \left(\ket{egg} + \ket{geg} - \ket{gge}\right)\otimes\ket{000},
	\label{eq:phasedwstate}
\end{equation}
where the phases are $\phi_1=\phi_2=0$, and $\phi_3=\pi$.
The order of the three phases is reflected in phases of the pump terms, where only the relative phases between the pumps eventually determine the excited state.
Choosing optimal parameters for the targeted generation yields a fidelity of about $83\,\%$.

\section{Equivalence of coherent qubit- and cavity driving schemes}
\label{appendix:pump_equivalence}
We begin by applying a unitary transformation which shifts the photon operators: $ a_i \mapsto b_i = a_i + \mu $, where $ \mu = \varepsilon_\mathrm{q}/g $.
The shifting does not change the bosonic commutation relations.
The transformed cavity and the interaction Hamiltonian have the form
\begin{align}
	H_\mathrm{q}^\prime = {} & \sum_{i=1}^{N} (\omega_\mathrm{q} - \omega_\mathrm{d}) \sigma_i^\dagger \sigma_i,\\
	\begin{split}
		H_\mathrm{c}^\prime = {} & \sum_{i=1}^{N} (\omega_\mathrm{c} - \omega_\mathrm{d}) b_i^\dagger b_i - \sum_{i,j=1}^{N} J_{ij} b_i^\dagger b_j \\
		& + \sum_{i=1}^{N} \mu \left(\sum_{j=1}^{N} J_{ij} - \omega_\mathrm{c} + \omega_\mathrm{d}\right) \left(b_i^\dagger + b_i\right),
	\end{split}\\
	H_\mathrm{c,q}^\prime = {} & \sum_{i=1}^{N} g \left(b_i^\dagger \sigma_i^{-} + b_i \sigma_i^{+}\right),
\end{align}
where we used $ J_{ij} = J_{ji} $ and neglected constant terms. The transformation eliminates the qubit-pump term and introduces a term containing $ (b_i^\dagger + b_i) $.
The latter describes coherent excitation of photons in the cavity modes. The Lindblad  dissipator describing cavity decay with the collapse operators $ \mathcal{D}[\sqrt{\kappa}a_i]\rho $ is transformed accordingly
\begin{equation}
	 \mathcal{D}[\sqrt{\kappa}a_i]\rho \mapsto \mathcal{D}[\sqrt{\kappa}b_i]\rho + \mu \frac{\kappa}{2} \left[b_i^\dagger - b_i, \rho\right].
\end{equation}
The second term contributes to the von unitary part $ -\mathrm{i}[H^\prime,\rho] $ of the dynamics, and has the form of coherent cavity pumping with amplitude $ \mathrm{i}\kappa\mu/2 $.
The transformed cavity Hamiltonian with the additional terms reads
\begin{equation}
	H_\mathrm{c}^\prime = \sum_{i=1}^{N} (\omega_\mathrm{c} - \omega_\mathrm{d}) b_i^\dagger b_i - \sum_{i,j=1}^{N} J_{ij} b_i^\dagger b_j + \sum_{i=1}^{N} \left(\varepsilon_\mathrm{c}^{i} b_i^\dagger + {\varepsilon_\mathrm{c}^{i}}^\ast b_i\right),
\end{equation}
where $ \varepsilon_\mathrm{c}^{i} = \mu (\sum_{j} J_{ij} - \omega_\mathrm{c} + \omega_\mathrm{d}) + \mathrm{i} \mu \kappa/2 $.
We obtain real excitation amplitude by applying the unitary transformation $ H \mapsto UHU^\dagger $ with $ U = \prod_{i} \exp(\mathrm{i}\varphi a_i^\dagger a_i) $.
As a result $ \varepsilon_\mathrm{c}^{i} \mapsto \varepsilon_\mathrm{c}^{i} \exp(\mathrm{i}\varphi) $, where the phase $ \varphi $ is chosen such that $ \varepsilon_\mathrm{c}^{i} $ becomes real.
We find for the cavity driving strength
\begin{equation}
	\varepsilon_\mathrm{c}^{i} = \frac{\varepsilon_\mathrm{q}}{g} \sqrt{\left(\sum_{j=1}^{N} J_{ij} - \omega_\mathrm{c} + \omega_\mathrm{d}\right)^2 + \left(\frac{\kappa}{2}\right)^2}.
	\label{eq:cavitypumpone}
\end{equation}
Coherently exciting the qubits is equivalent to individual excitation of the cavities, although at modified amplitudes $\varepsilon_\mathrm{c}^{i}$.
In the case where all cavity couplings are equal, i.e.~$J_{ij} \equiv J$, the cavity driving amplitudes in Eq.~\eqref{eq:cavitypumpone} simplify to
\begin{equation}
	\varepsilon_\mathrm{c}^{i} \equiv \varepsilon_\mathrm{c} = \frac{\varepsilon_\mathrm{q}}{g} \sqrt{\left[\left(N-1\right)J - \omega_\mathrm{c} + \omega_\mathrm{d}\right]^2 + \left(\frac{\kappa}{2}\right)^2}.
	\label{eq:cavitypumptwo}
\end{equation}

\section{Derivation of rate equations from the Bloch-Redfield equation}
\label{appendix:rates}
We divide the Bloch-Redfield equation
\begin{equation}
	\begin{aligned}
		\frac{\mathrm{d}}{\mathrm{d} t}\rho(t) = {} & \underbrace{- \mathrm{i} [H, \rho(t)]}_{\substack{\mathrm{I}}} + \underbrace{\sum_{\alpha,\omega} \gamma_\alpha (\omega) A_\alpha(\omega) \rho(t) A_\alpha^\dagger (\omega)}_{\substack{\mathrm{II}}} \\
		& - \underbrace{\frac{1}{2} \sum_{\alpha,\omega} \gamma_\alpha(\omega) \{ A_\alpha^\dagger(\omega) A_\alpha(\omega),\rho(t) \} }_{\substack{\mathrm{III}}}
	\end{aligned}
\end{equation}
into three different parts and switch into the Hamiltonian eigenbasis
\begin{equation}
	H \ket{k} = E_k \ket{k},
\end{equation}
where the operators have the form
\begin{align}
	\rho &= \sum_{kl} \rho_{kl} \ket{k}\bra{l}, \\
	H &= \sum_k E_k \ket{k}\bra{k}, \\
	A_\alpha(\omega_{kl}) &= A_{lk}^\alpha \ket{l}\bra{k}.
\end{align}
With this, the first part simplifies to
\begin{equation}
	\mathrm{I} = - \mathrm{i} \sum_{kl}(E_k - E_l) \rho_{kl}\ket{k}\bra{l},
\end{equation}
representing a coupling of each off-diagonal element to itself and resulting in an oscillation with frequency $\omega_{kl} = E_k - E_l$.
Thus, part I does not influence the diagonal elements, since the energy difference is zero.
For the second and third part we note that the sum over all transition frequencies $\omega$ is equal to the sum over all $k$ and $l$ by replacing $\omega$ with $\omega_{kl}$.
The second part then simplifies to
\begin{equation}
	\mathrm{II} = \sum_{\alpha, kl} \gamma_\alpha(\omega_{kl})\rho_{kk}|A^\alpha_{kl}|^2\ket{l}\bra{l},
\end{equation}
coupling the diagonal elements with each other, but all terms containing off-diagonal elements vanish.
The third term simplifies to
\begin{align}
	\mathrm{III} &= -\frac{1}{2} \sum_{\alpha, kli}\left( \gamma_\alpha(\omega_{ki})|A^\alpha_{ki}|^2 + \gamma_\alpha(\omega_{li})|A^\alpha_{lm}|^2 \right)\rho_{kl} \ket{k}\bra{l} \nonumber\\
	&\equiv - \sum_{\alpha, kl}\lambda^\alpha_{kl} \rho_{kl}\ket{k}\bra{l}.
	\label{eq:part3}
\end{align}
Here, the matrix elements also only couple to themselves.
The parameter 
\begin{equation}
	\lambda^\alpha_{kl} = \frac{1}{2} \sum_{i}\left( \gamma_\alpha(\omega_{ki})|A^\alpha_{ki}|^2 + \gamma_\alpha(\omega_{li})|A^\alpha_{li}|^2 \right)
\end{equation}
is positive, so that Eq.~\eqref{eq:part3} describes a decay of the matrix elements.
For $k = l$, it simplifies to
\begin{equation}
	\lambda^\alpha_{kk} = \sum_{i}\gamma_\alpha(\omega_{ki})|A^\alpha_{ki}|^2.
\end{equation}
We can combine these results to obtain a set of linear differential equations
\begin{align}
	\dot{\rho}_{kl} &= - \left[ \sum_\alpha \lambda^\alpha_{kl} + \mathrm{i} \omega_{kl} \right] \rho_{kl} \quad \mathrm{for} \quad k\neq l \\
	\dot{\rho}_{kk} &= \sum_{\alpha,i} \left[ \gamma_\alpha(\omega_{ik}) |A^\alpha_{ik}|^2 \rho_{ii} -\gamma_\alpha(\omega_{ki})|A^\alpha_{ki}|^2 \rho_{kk} \right].
\end{align}
The off-diagonal elements only couple to themselves, and the solution is an exponential decay at rate $\lambda$ combined with an oscillation with the respective frequency.
The steady-state $\rho_\mathrm{ss} = \rho(t \rightarrow \infty)$ will thus be diagonal in the Hamiltonian's eigenbasis.
The equation for the diagonal elements can be written as a set of rate equations
\begin{equation}
	\dot{\rho}_{kk} = \sum_{i} \left[ \Gamma_{i \rightarrow k} \, \rho_{ii} - \Gamma_{k \rightarrow i} \, \rho_{kk} \right]
\end{equation}
with the transition rates
\begin{equation}
	\Gamma_{i \rightarrow k} = \sum_\alpha \gamma_\alpha(\omega_{ik}) |A^\alpha_{ik}|^2.
\end{equation}

\section{Derivation of effective Hamiltonian using the Schrieffer-Wolff transformation}
\label{appendix:approximations}
The following derivation follows the concepts introduced in Ref.~\cite{aron_steady-state_2014}.
We apply the Schrieffer-Wolff transformation to the Hamiltonian in Eq.~\eqref{eq:hamiltonian_general} to achieve a decoupling of the qubit and cavity sector of the Hilbert space up to second order in $g/\Delta$. For this, we use a slightly different partition for the Hamiltonian
\begin{equation}
	H = H_\mathrm{q} + H_\mathrm{c} + H_\mathrm{c,q} + H_\mathrm{\varepsilon},
\end{equation}
where $H_\mathrm{\varepsilon}$ does describe the cavity drive, which is \textbf{not} included in $H_\mathrm{c}$.
In the rotating wave approximation, the Hamiltonians  read:
\begin{align}
	H_\mathrm{q} &= \sum_{i=1}^{2} (\omega_\mathrm{q}-\omega_\mathrm{d}) \sigma_i^{+} \sigma_i^{-},\\
	H_\mathrm{c} &= \sum_{i=1}^{2} (\omega_\mathrm{c}-\omega_\mathrm{d}) a_i^\dagger a_i - \sum_{i,j=1}^{2} J_{ij} a_i^\dagger a_j,\\
	H_\mathrm{c,q} &= \sum_{i=1}^{2} g \left(a_i^\dagger \sigma_i^{-} + a_i \sigma_i^{+}\right),\\
	H_\mathrm{\varepsilon} &= \sum_{i=1}^{2} \varepsilon_{\mathrm{c}}\left( a_i^\dagger + a_i \right).
\end{align}
To further simplify the system, we introduce symmetric 
\begin{align}
	A &= (a_1 + a_2)/\sqrt{2} \\
	\Sigma^\pm &= (\sigma_1^\pm + \sigma_2^\pm)/\sqrt{2}
\end{align}
and antisymmetric operators
\begin{align}
	a &= (a_1 - a_2)/\sqrt{2} \\
	\sigma^\pm &= (\sigma_1^\pm - \sigma_2^\pm)/\sqrt{2}.
\end{align}
The photonic modes $A$ and $a$ diagonalize $H_\mathrm{c}$ with eigenenergies $\omega_\mathrm{c}^\mp$.
In this basis, the Hamiltonians read
\begin{align}
	H_\mathrm{q} &= (\omega_\mathrm{q}-\omega_\mathrm{d}) \left[ \Sigma^+\Sigma^- + \sigma^+\sigma^- \right],\\
	H_\mathrm{c} &= (\omega_\mathrm{c} - J -\omega_\mathrm{d}) A^\dag A + (\omega_\mathrm{c} + J -\omega_\mathrm{d}) a^\dag a,\\
	H_\mathrm{c,q} &= g \left(A^\dagger \Sigma^- + A\Sigma^+ + a^\dagger \sigma^- + a\sigma^+ \right),\\
	H_\mathrm{\varepsilon} &= \varepsilon_{\mathrm{c}} (A^\dagger + A) / \sqrt{2}.
\end{align}
We can eliminate the qubit-cavity-coupling in first order in $g/\Delta$ by applying the Schrieffer-Wolff transformation, $H \mapsto \mathrm{e}^X H \mathrm{e}^{-X}$, with the generator
\begin{equation}
	X = g \left[ \frac{A\Sigma^+ - A^\dagger\Sigma^-}{\omega_\mathrm{q} - \omega_\mathrm{c}^-} + \frac{a\sigma^+ + a^\dagger \sigma^-}{\omega_\mathrm{q} - \omega_\mathrm{c}^+} \right].
\end{equation}
It fulfills the condition $[X, H_\mathrm{q} + H_\mathrm{c}] = H_\mathrm{c,q}$, implying that we do a Schrieffer-Wolff in terms of the undriven Hamiltonian.
This is possible since the term $[X, H_\mathrm{\varepsilon}]$, which is first order in $g/\Delta$, does not couple the qubit- and cavity subspaces, and we do not acquire a first order coupling through this additional term.
Truncating third and higher orders in $g/\Delta$ and applying the mean-field approximation by decomposing the photon fields into mean field plus fluctuations
\begin{align}
	A &\equiv \bar{A} + D, & a &\equiv \bar{a} + d \\
	\bar{A} &= \frac{\sqrt{2}\varepsilon_\mathrm{c}}{\omega_\mathrm{d} - \omega^-_\mathrm{c} + \mathrm{i} \kappa/2}, & \bar{a} &= 0
\end{align}
and neglecting quadratic terms of the form $g^2D^\dagger D$, we arrive at the transformed Hamiltonian
\begin{equation}
	\tilde{H} = \tilde{H}_\mathrm{q} + \tilde{H}_\mathrm{c} + \tilde{H}_\mathrm{c,q}.
\end{equation}
The qubit-Hamiltonian
\begin{equation}
	\tilde{H}_\mathrm{q} = \sum_{i=1}^2 \left( \frac{\Omega_\mathrm{R}}{2}\sigma^x_i + \frac{\tilde{\Delta}}{2}\sigma^z_i \right) - \frac{1}{2}J\frac{g^2}{\Delta^2} [\sigma^x_1\sigma^x_2 + \sigma^y_1\sigma^y_2]
\end{equation}
now has renormalized energies
\begin{equation}
	\begin{aligned}
		\tilde{\Delta} = {} & \omega_\mathrm{q} - \omega_\mathrm{d} + \left( \frac{g}{\Delta + J} \right)^2 \bigg( (\Delta+J) |\bar{A}|^2 \\\
		& + \frac{(\Delta + J)^2}{\Delta} +\sqrt{2} \varepsilon_\mathrm{c} \mathrm{Re}\bar{A} \bigg),
	\end{aligned}
	\label{eq:renormqenergies}
\end{equation}
a single-qubit $\sigma^x$ term $\Omega_\mathrm{R} = 2g\varepsilon_\mathrm{c}/(\Delta + J)$ caused by the drive, and an effective qubit-qubit interaction that is second order in $g/\Delta$.
The photonic part is still described by the two independent modes
\begin{equation}
	\tilde{H}_\mathrm{c} = (\omega_\mathrm{c} -\omega_\mathrm{d} - J)D^\dagger D +(\omega_\mathrm{c} -\omega_\mathrm{d} + J)d^\dagger d.
\end{equation}
The interaction between the two parts is now given by
\begin{equation}
	\begin{aligned}
	\tilde{H}_\mathrm{c,q} = {} & \frac{1}{2} \left( \frac{g}{\Delta} \right)^2
	 \left[ \left( \bar{A}\Delta + \frac{\varepsilon_\mathrm{c}}{\sqrt{2}} \right)D^\dagger (\sigma^z_1 + \sigma^z_2) \right.
	\\
	& + \left.\left( \bar{A}\Delta + \frac{\varepsilon_\mathrm{c}}{\sqrt{2}} \right)d^\dagger (\sigma^z_1 - \sigma^z_2)\right] + \mathrm{h.c.\,}
	\end{aligned}
\end{equation}
and is also second order in $g/\Delta$, meaning this effective interaction is weaker than before the Schrieffer-Wolff transformation.
It is of the form used in Eq.~\eqref{eq:h_cq_tilde}, where
\begin{equation}
	c = \frac{1}{2} \left( \frac{g}{\Delta} \right)^2 \left( \bar{A}\Delta + \frac{\varepsilon_\mathrm{c}}{\sqrt{2}} \right).
\end{equation}

\bibliography{references}

\end{document}